# Stacking dependence of carrier transport properties in multilayered black phosphorous

A. Sengupta[1,2,3*], M. Audiffred[3], T. Heine[3,4], and T. A. Niehaus[2]

[1]*Advanced Semiconductors and Computational Nanoelectronics Lab, School of VLSI Technology, Indian Institute of Engineering Science and Technology, Shibpur, Howrah – 711 103, India*
[2]*Institute-I Theoretical Physics, Universität Regensburg, 93040 Regensburg, Germany*
[3]*Department of Physics and Earth Science, Jacobs University, Bremen, Campus Ring 1, 28759 Bremen, Germany*
[4]*Wilhelm-Ostwald-Institut für Physikalische und Theoretische Chemie, Universität Leipzig, Linnéstr. 2, 04103 Leipzig, Germany*

*(\*Corresponding Author e-mail: a.sengupta@vlsi.iiests.ac.in, Fax: +91-33- 2668-2916)*

***Abstract:*** *We present the effect of different stacking orders on carrier transport properties of multi-layer black phosphorous. We consider three different stacking orders AAA, ABA and ACA, with increasing number of layers (from 2 to 6 layers). We employ a hierarchical approach in density functional theory (DFT), with structural simulations performed with Generalized Gradient Approximation (GGA) and the bandstructure, carrier effective masses and optical properties evaluated with the Meta-Generalized Gradient Approximation (MGGA). The carrier transmission in the various black phosphorous sheets was carried out with the non-equilibrium Green's function (NEGF) approach. The results show that ACA stacking has the highest electron and hole transmission probabilities. The results show tunability for a wide range of band-gaps, carrier effective masses and transmission with a great promise for lattice engineering (stacking order and layers) in black phosphorous.*

## 1. Introduction

In recent years non-graphitic layered materials such as transition metal dichalcogenides and black phosphorous have generated a lot of interest as potential candidates for post silicon electronic devices. The layered nature of these materials and advances in nanofabrication techniques have made it possible to realize MOSFETs based on single or multilayered $MoS_2$ [1,2] and more recently few layered black phosphorous. [3,4] In particular, black phosphorous has been found to be particularly promising as a candidate for p-type FETs. [4]

Black phosphorous, when exfoliated to single or few layer phosphorene, can show significant carrier anisotropy depending upon the number of layers involved and other factors such as strain. [5-16] Recently, the stacking order of the individual layers in black phosphorous has also shown to create variations in the energy band-gap and optical properties. [7,11,13] From the point of bottom-up nanofabrication, tailoring the electrical properties of black phosphorous through layer engineering [17] (i.e. arrangement of the stacking order in layered sheets) could be of interest for controlling the carrier transport properties of such multilayer 2-D phosphorene. In this paper we present a computational study of the electronic properties of multilayer (2–6 layers) black phosphorous sheets under different stacking configurations. Three different stacking orders namely AAA, ABA and ACA (Fig. 1), were considered for studying their electronic properties and carrier transport behavior.

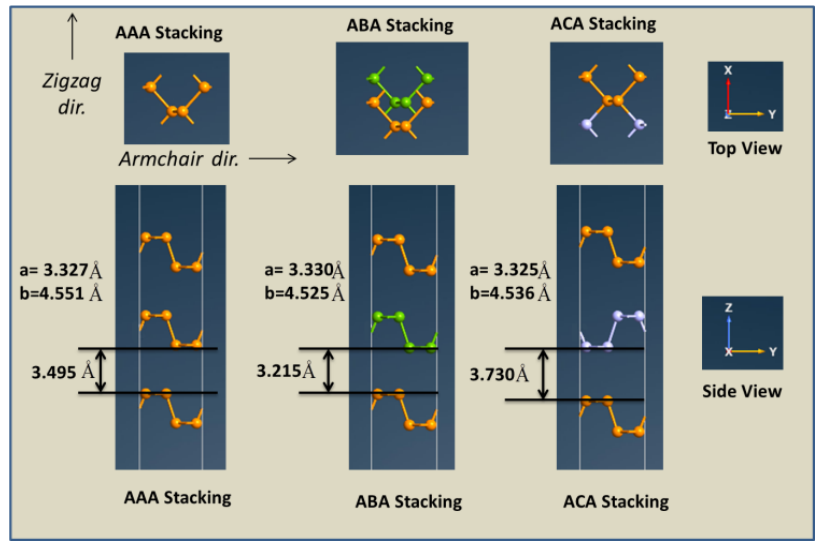

**Fig. 1.** The optimized structures for layered black phosphorous showing the side and top views of the stacked unit-cell. Layers in site A , B and C are color coded for better understanding of the orientation from top.

Properties such as bandstructure, interlayer interaction energy, electron and hole effective masses, and optical properties have been calculated by density functional theory (DFT) while the carrier transport properties in the differently stacked layered black phosphorous sheets are evaluated with DFT based - Non-equilibrium Green's function (NEGF) method.

## 2. Methodology

An important and tricky issue for the simulation of 2D black phosphorous is the choice of correct basis and exchange correlation function in DFT. [3-9], Local Density Approximation (LDA) and Generalized Gradient Approximation (GGA) mostly suffer from a large underestimation of the energy band-gaps, which is particularly evident in case of phosphorous. [3-9] Hybrid methods such as Heyd-Scuseria-Ernzerhof (HSE) [18,19] produce much more accurate band-gaps than Kohn-Sham DFT, [20] but these are not suited for accurate estimation of carrier effective masses [8] and thus not preferable for carrier transport calculations. The problem with HSE in effective mass estimation is described by Cai et. al. [8] to originate from the Hartree-Fock exchange, that replaces the semi-local exchange correlation functional in HSE. [18] Methods such as meta- Generalized Gradient Approximation (MGGA) as implemented by Tran and Blaha (TB09LDA) [21,22] can produce good accuracy in band-gaps (comparable to HSE) with an improvement in the effective mass estimation over HSE (as the exchange correlation used in MGGA is LDA based). [22-24] In this context MGGA can be a suitable method for simulation of transport in black phosphorous as it produces fairly accurate band gaps, along with good estimation of the effective mass. The GLLB-SC model is also sometimes used as an alternative approach to predict the band gap with comparable quality to those determined at the GW level of theory [25]

However for the geometrical structure optimization in TB09LDA Meta-GGA has issues due to difficulties in defining a proper derivative (necessary for force calculation) of the exchange correlation potential. [21-24] For the geometric structure optimization which is necessary for accurate description of multi-layer systems, LDA or GGA remains the better option. [23- 27] For the correction due to the interlayer van der Waals forces in layered structures in DFT, Grimme's DFT-D2 method can be suitable. [24,28] Thus to attack the problem at hand i.e. simulation of carrier transport in multilayer black phosphorous, a hierarchical approach is adopted using GGA for structure optimization and MGGA for simulation of electronic and optical properties.

## 2.1. Structure Optimization and Formation Energy Calculations

In our work, we use the QuantumWise AtomistixToolKit (ATK) 14.2 for materials simulations. For the structure optimization and interlayer interaction energy estimation, we employ GGA in DFT with a Perdew-Burke-Ernzerhof (PBE) exchange correlation functional [27] with a Double Zeta Polarized (DZP) basis set. [21,24] GGA was employed owing to its superiority over hybrid pseudopotentials for force and energy optimization. We employ a Troullier-Martins type norm-conserving pseudopotential set (FHI [Z=5] DZP for Phosphorous). Relativistic core corrections are included in the pseudopotential. [24] A 9x9x2 Monkhorst-Pack [29] k-grid is used for the simulations. A cutoff energy of 180 Rydberg is considered with Pulay mixer algorithm [30] of 0.0002 Rydberg tolerance and 100 maximum steps. Grimme's DFT-D2 correction [24,28] is employed to account for the van der Waals interactions. The parameters used in Grimme's DFT-D2 for phosphorous are scale factor of 0.75, damping factor of 20 and cutoff distance of 30 $\text{Å}$. The $R_0$ and the $C_6$ parameters for phosphorous are $1.7050\,\text{Å}$ and 7.84 J nm$^6$/ mol respectively. The optimization is performed with a limited memory Broyden-Fletcher-Goldfarb-Shanno (LBFGS) algorithm [24,31] The formation energy per atom is calculated by the total energy calculations in the GGA-PBE simulations.

The optimized structures for the multilayer black phosphorous are shown in Fig. 1. The A-A, A-B and A-C interlayer spacing and the unit-cell parameters are also presented in the Fig. 1. In all the structures a vacuum of 20 Å on both top and bottom in c-direction is maintained. The lattice parameters and interlayer spacing obtained in our simulations are in agreement with those reported by Dai et. al. [7] and Çakir et. al. [13] for bilayer phophorene. These structures are used in further simulations for bandstructure and carrier transport properties with MGGA-TB09LDA. The formation energy per atom is defined as

$$\Delta E_{form} = \frac{E_n}{N_n} - \frac{E_1}{N_1} \qquad (1)$$

$n$ is the number of layers, $E_n$ the total energy for the n-layered system and $E_1$ the total energy of a single layer of black phosphorous, $N_1$ and $N_n$ signify the total number of atoms in unit cell of the single layered and the $n$ layered system.

## *2.2. Simulation of electronic properties and transport*

For the bandstructure, optical properties and the transport calculations we employ MGGA-TB09LDA exchange correlation function, as it offers better band-gap estimation for black phosphorous than LDA or GGA-PBE [3-9] and at the same time superior effective mass estimation compared to HSE06 [8] The other computational details are as for structure and energy calculations as specified in 2.1.

The effective mass is evaluated from the energy band-structure using the parabolic fitting formula [13] (in the effective mass analyzer of ATK).

$$m^*_{i,j} = \hbar^2 \left( \frac{\partial^2 E}{\partial k_i \partial k_j} \right)^{-1} \qquad (2)$$

$k_{i,j}$ is the wave-vector along the $i(j)$ direction and $E$ is the energy eigenvalue. The masses are evaluated along the armchair (zigzag) edges w.r.t. the unit cell axes. The $k$ mesh for effective mass evaluation uses 15 points in the $i(j)$ direction with a separation of 0.001 Å$^{-1}$.

The transmission calculations are realized in ATK following the Landauer formalism. [32-36] In the NEGF method the-Poisson Schrödinger equation of the system is solved self-consistently. Setting up the self-energy matrices $\Sigma_1$ and $\Sigma_2$ for the infinite contacts, the Green's function $G$ is constructed as

$$G(E) = \left[ EI - H - \Sigma_1 - \Sigma_2 \right]^{-1} \qquad (3)$$

In (2) $I$ is the identity matrix. From (2) parameters like the broadening matrices $\Gamma_1$ and $\Gamma_2$ and the spectral densities $A_1$ and $A_2$ (of the two contacts) are evaluated as [32-36]

$$\Gamma_{1,2} = i\left[\Sigma_{1,2} - \Sigma_{1,2}^\dagger\right] \quad (4)$$

$$A_{1,2}(E) = G(E)\Gamma_{1,2}G^\dagger(E) \quad (5)$$

For DFT-NEGF in ATK, a multi-grid self-consistent Poisson-Schrödinger solver is employed. The converged values are used to evaluate the transmission matrix $\Im(E)$ as

$$\Im(E) = trace\ [A_1\Gamma_2] = trace\ [A_2\Gamma_1] \quad (6)$$

For the transmission spectra we use the Krylov self-energy calculator [24, 32, 33] with the average Fermi level being set as the energy zero parameter.

For the evaluation of the optical properties, the Kubo-Greenwood formula, given below [24,37] is employed to evaluate the susceptibility tensor.

$$\chi_{ij}(\omega) = -\frac{e^2\hbar^4}{m_e^2\varepsilon_0 V\omega^2}\sum_{nm}\frac{f(E_m)-f(E_n)}{E_{nm}-\hbar\omega-i\hbar\Gamma}\Pi^i_{nm}\Pi^j_{mn} \quad (7)$$

In this formula the dipole matrix element between the $m$-th and $n$-th state is represented by $\Pi^i_{nm}$. $m_e$ is electron mass, $V$ volume, $\Gamma$ is broadening and $f(.)$ is the Fermi function. From this the complex dielectric constant can be found as

$$\epsilon_r(\omega) = (1 - \chi(\omega)) \quad (8)$$

The refractive index $n$ is related to $\epsilon_r$ as

$$n + i\kappa = \sqrt{\epsilon_r} \quad (9)$$

$\kappa$ being the extinction coefficient. [24, 37] With a knowledge of $\kappa$, the optical absorption coefficient $\alpha$ can be calculated by [24, 37]

$$\alpha = 2\frac{\omega}{c}\kappa \quad (10)$$

In our case we study the perpendicular component of the optical absorption coefficient as

$$\alpha_\perp = \frac{1}{2}(\alpha_{xx} + \alpha_{yy}) \quad (11)$$

The electron-phonon couplings for the three different stacking types are also calculated. For the phonon calculations involved in this we use a supercell based small displacement method also referred to as *frozen phonon model*. [24,38] For phonon calculations we use a 9x9x1 supercell. Due to the large number of atoms and displacements involved, the phonon calculations were carried out using the Slater-Koster method (with the DFTB-CP2K parameters available in ATK) [39]. The dynamical matrix elements are calculated using the acoustic sum rule and a central finite difference method applying 0.01 Å displacements to the supercell under consideration in the x and y directions. A (15,1,1) k-mesh and a (15,15,1) q-mesh was used for the electron – phonon coupling calculations. [24] The electron-phonon coupling strength under adiabatic approximation is given as [38]

$$g_{kq}^{\lambda} = \sqrt{\frac{\hbar}{2mN_C\omega_{q\lambda}}} \cdot M_{kq}^{\lambda} \qquad (12)$$

In (11), $q$ is the wave vector of the phonon mode and its branch index is $\lambda$, $k$ is the carrier (electron/hole) wave vector. The associated phonon frequency is $\omega_{k\lambda}$, $m$ is the carrier effective mass, $N_C$ the number of unit cells involved and $M_{kq}^{\lambda}$ is the coupling matrix element. [24,38]

## 3. Results & Discussion

In Fig. 2, we show the bandstructure of monolayer black phosphorous simulated with different functional such as LDA-PZ [25], GGA-PBE [27], the GLLB-SC model [25] and the MGGA-TB09LDA [21]. The k-points used in these cases were 9x9x1 and cut-off energy was set to 180 Rydberg.

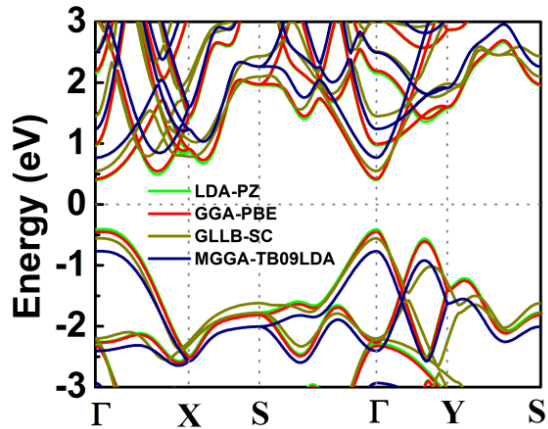

**Fig. 2.** The bandstructure of monolayer black phosphorous calculated with different functionals.

As expected for LDA and GGA we observe a significant band-gap under-estimation compared to experimental [5] values (~1.46 eV) for black phosphorous, with LDA predicting a band gap of 0.82 eV and GGA giving the value 0.89 eV. GLLB-SC provides a somewhat better estimation of band gap value of about 1.15 eV. MGGA is found to give similar value of band-gap (1.52 eV) as also predicted with HSE06 results of other groups. [7,8]

In the study of the formation energy per atom (Fig. 3) of the various layered stacks we observe that ABA stacking has the most stable configuration, having the least amount of formation energy per atom among the three, followed by the AAA and the ACA stackings. The ABA stacks have ~0.03 eV lesser formation energy/ atom compared to ACA and ~0.015 eV less than AAA stacked black phosphorous. Similar formation energies (with DFT-D2 / DFT-vdW) for bilayer black phosphorous has also been reported by Shulenburger et. al. [12] and Çakir et. al. [13]. With the increasing number of layers, there tends to be a very slight change in the formation energy per atom. Although these results indicate that naturally black phosphorus multilayers are more likely to stack up in ABA order, it is also possible to stack engineer AAA and ACA phosphorous, as the formation energies differ in magnitude by few meV only.

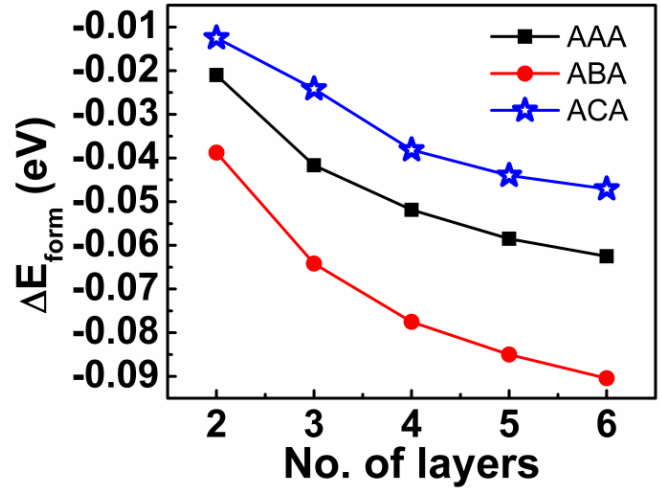

**Fig. 3.** The formation energy per atom (obtained with GGA-PBE) of various stacked multilayer black phosphorous.

It is worth mentioning that the interlayer interaction cannot be fully described by DFT-D2 as this interaction is not purely dispersive, as reported by other groups. [12, 14] Nevertheless, it gives a qualitatively similar trend as some other methods which treat electron correlation at a more advanced level (e.g., *Quantum Monte-Carlo*). [12]

As we further study the layered stacks for their bandstructure and electronic properties, it is observed (in Figs. 4 and 5) that with the addition of layers the band gaps tends to close, however the nature of the gap remains direct for the bilayered (2L) to the six-layered (6L) black phosphorous, irrespective of the nature of stacking. Overall the AAA and ABA stacked black phosphorous have higher band-gaps compared to the ACA stacked phosphorous, with AAA showing the biggest band gap.

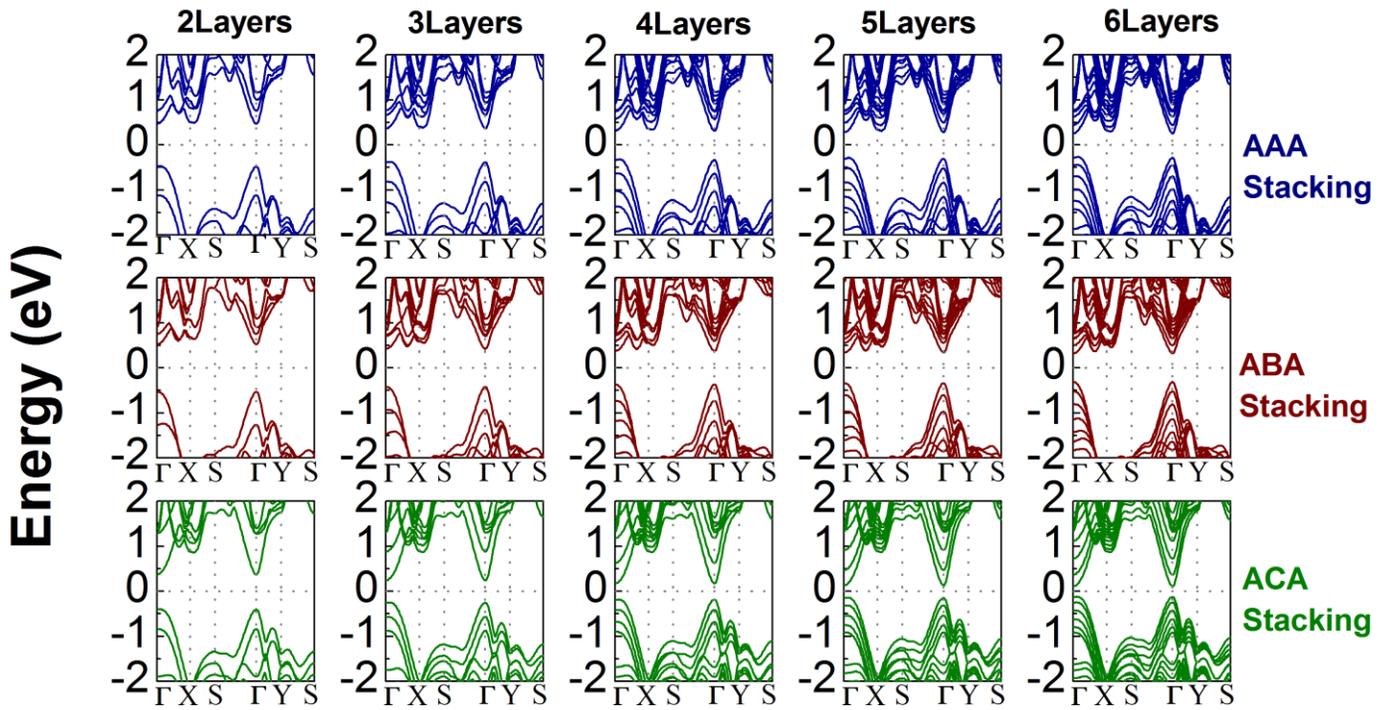

**Fig. 4.** The bandstructures of various stacked black phosphorous calculated with MGGA-TB09LDA.

The rate of band gap closing for the ACA stacking seems somewhat faster than that for AAA and ABA when increasing the number of layers. In this regard the rate of upward shift of the valence bands (VB) is faster than that for the downward shift of conduction bands (CB) in AAA and to some extent also in ABA stacking. In ACA stacking however the closing of the gap on the CB side seems faster than that on the VB side. The bilayer (2L) ABA displays the largest band-gap of about 1.06 eV followed by 2L AAA with 0.956 eV and 2L ACA 0.784 eV. The band-gap values for the bilayer systems match with those calculated with HSE06 by Dai et. al. [7] and Çakir et. al. [13]. As the number of layers is added the band gap reduces to 0.644 eV for 6L ABA , 0.55 eV for 6L AAA and 0.244 eV for 6L ACA stacked multilayers. Thus a high degree of tunability in terms of band gap is achievable with controlling the number of layers and the stacking order in layered black phosphorous. When we extend the calculations to an infinite number of layers, the gaps reduce down to 0.47, 0.54 and 0.052 eV for AAA, ABA and ACA stacking respectively.

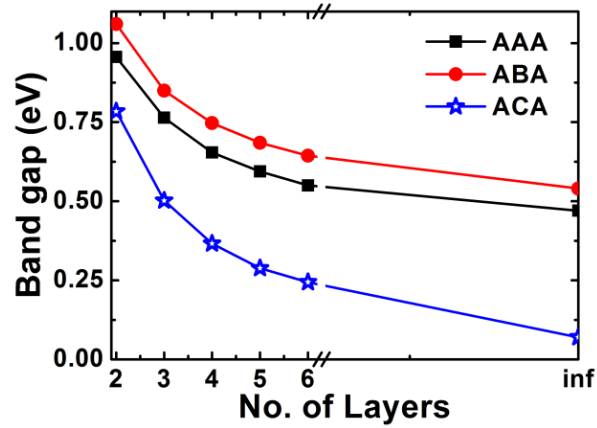

**Fig. 5.** The variation of band-gap (calculated with MGGA-TB09LDA) of black phosphorus with the nature of stacking and the number of layers.

This range of variation of band-gap for black phosphorous is wider than layered $MoS_2$ where LDA-PZ results show band-gap of 1.23 eV for bilayer and 0.82 for bulk $MoS_2$. Also the nature of the band gap is maintained (direct) in case of layered black phosphorous, whereas for $MoS_2$ and other transition metal dichalcogenides, there is a transition from direct to indirect band-gap as soon as the second layer is added. The retention of the direct band-gap nature in all the cases of layered black phosphorous makes it more promising in terms of applicability for tunnel FET devices.

The calculated effective masses for electrons and holes at the conduction band minima ($\Gamma$ point in all the cases) and valence band maximum (also $\Gamma$ point) respectively, in the armchair and the zigzag directions are shown in Fig. 6. We observe that for all the stacking orders there is a considerable difference in carrier effective masses in the armchair and the zigzag directions. For the zigzag direction, ACA stacking has the smallest effective masses for both holes and electrons, and the value of the carrier effective mass for such stacking does not change significantly with the addition of layers. ABA stacking shows the highest carrier effective masses from 2L through to 6L stacks under study. For ABA there exists a slight reduction of masses with the addition of layers, as compared to that for the ACA stacking.

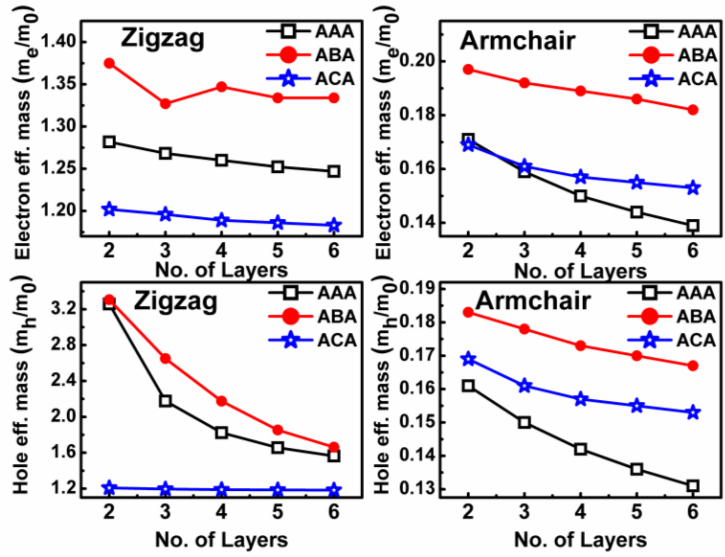

**Fig. 6.** The electron and hole effective masses at Γ point in the armchair and the zigzag directions for the differently stacked multilayer black phosphorous. (calculated with MGGA-TB09LDA)

The reduction in effective masses, with the increasing number of layers is more significant in case of holes than electrons in general. In the zigzag direction though, the hole effective mass seems to vary little with increasing layers for the ACA stacking.

Coming to the armchair direction, the effective mass for ACA stacking is intermediate between ABA (highest) and AAA (least). Here though the reduction in electron / hole masses with increase in number of layers is slightly less rapid in ACA as in AAA and ABA.

If we focus on the wide range of effective masses possible to obtain with layer engineering in black phosphorous, we observe a variation in the electron effective mass from 1.38 (2L-ABA in zigzag direction) to 0.14 $m_0$ (6L-AAA in armchair direction) . For holes this variation is 3.2 (2L-ABA in zigzag direction) to 0.13 $m_0$ (6L-AAA in armchair direction). In comparison $MoS_2$ effective masses for electron vary only from 0.612 (bilayer $MoS_2$) to 0.57 $m_0$ (bulk $MoS_2$), and the hole mass varies from 1.02 (bilayer) to 0.668 $m_0$ (bulk $MoS_2$). Overall this sort of effective mass variations in black phosphorous suggest a high degree of tunability with stacking order, selection of transport direction and also the number of layers (especially for hole effective masses in AAA and ABA stacked black phosphorous).

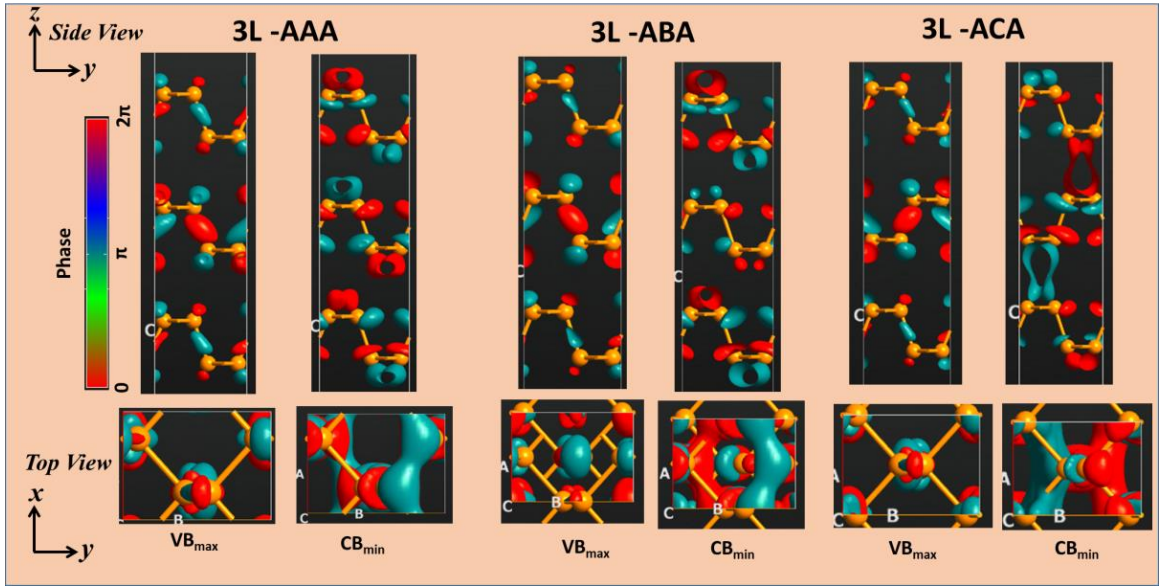

**Fig. 7:** Spatial structure of wavefunctions of the VBM and the CBM states illustrated in the xy (top view) and yz (side view) planes using an isosurface of 0.06 e Å$^{-3}$.

In Fig. 7 we show the wavefunctions of the states at valence band maxima (VBM) and conduction band minima (CBM) for the 3 types of stacking. The trilayer stacks are chosen to show the nature of interlayer interactions clearly. In ACA stacking the very prominent bonding states are observed between the crest and trough of the puckered honeycombs of the individual layers. Similar bonding states, though less prominent are also observed between the layers of ABA and AAA stacking. The observations for AA and AB stackings are consistent with the findings of Qiao et. al. [16]

From the simulations, the interlayer interaction thus appears the strongest in the case of ACA stacking. With the addition of layers the difference in interlayer interaction among the layers affects the bandstructure more and more. [13] The effect of interlayer bonding in ACA, only increases with the addition of more layers, and is responsible for the faster closing of the band gap in this type of stacking.

The charge densities for the 3 different stackings are also discussed in detail in the recent work by Çakir et. al. [13] Also, in another work by Wang et. al. [15] it is explained how the density changes from one type of stacking to the other. From these we are further able to imply that the interlayer interaction is not purely dispersive in nature.

We present the carrier transmission spectra in such layered sheets, in Fig. 8(a)-8.(b). In the Fig. 8 (a) & (b) the conductance has been scaled with respect to the width of the respective unit-cells. We observe that for the armchair direction the transmission channels both above and below the Fermi level ($E$=0), which signify the electron and hole conduction respectively, are both more or less equally transparent. Among the differently stacked layers, AAA stacking seems to have a more staircase-like profile (especially near the Fermi level) in its transmission spectra. This staircase nature seems less predominant with ABA or ACA stacks, and also with the addition of subsequent layers, the transmission spectra becomes more and more smeared. A probable explanation of this behavior lies in the AAA stacking being more symmetric / isotropic lattice than the ABA or the ACA type stacks. With increase in the number of layers, the number of conducting channels is incremented significantly and therefore the transmission also increases.

For the zigzag direction it is clearly seen that the hole transport is strongly suppressed in comparison to the electron transport. This stems from the asymmetry in the electron and hole effective masses (holes being significantly heavier) as evident in Fig. 6. The hole effective mass strongly depends on the type of stacking. The variations in the interlayer interaction strengths (owing to difference in stacking) causes flattening / sharpening of valence band maxima (VBM) and conduction band minima (CBM) in multilayer black phosphorous, thereby creating a disparity in effective masses.

In general, the effective masses are smaller in the armchair direction owing to strong $p_z$ contributions to the VBM and CBM. The interlayer bonding states can provide a suitable pathway for charge transport in these systems. If we again closely observe Fig. 7, (from the top views) it is evident that the isosurfaces belonging to these bonding states tend to overlap from one side of the sheet to another (considering periodicity in the $xy$ plane), which corresponds to the armchair / zigzag directions. These interlayer bonding states being more conducive to charge transport where the overlap is larger (e.g. for ACA) the stacks as a whole become more conducting.

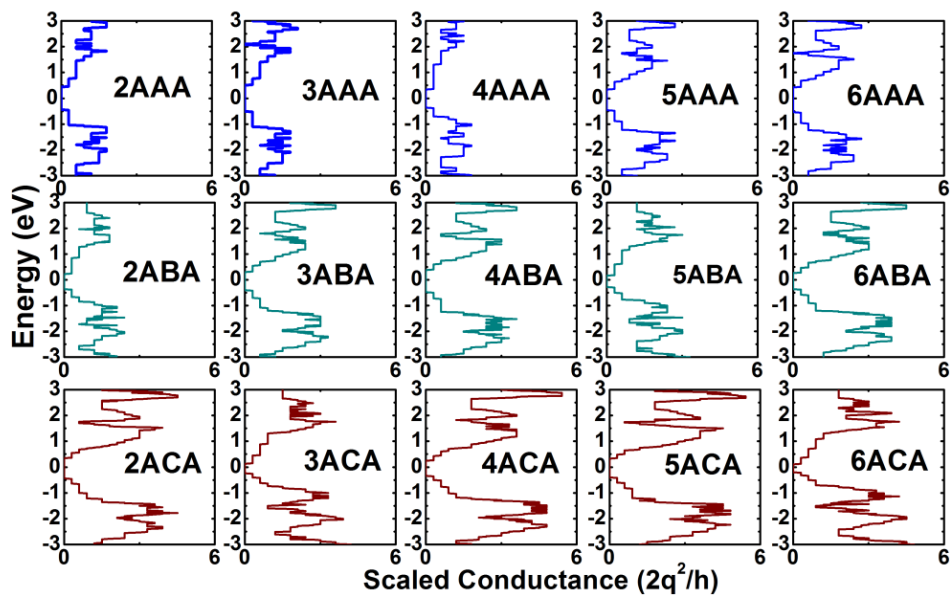

**Fig. 8.(a)**The scaled transmission spectra in armchair direction for different stacking of layered black phosphorous.

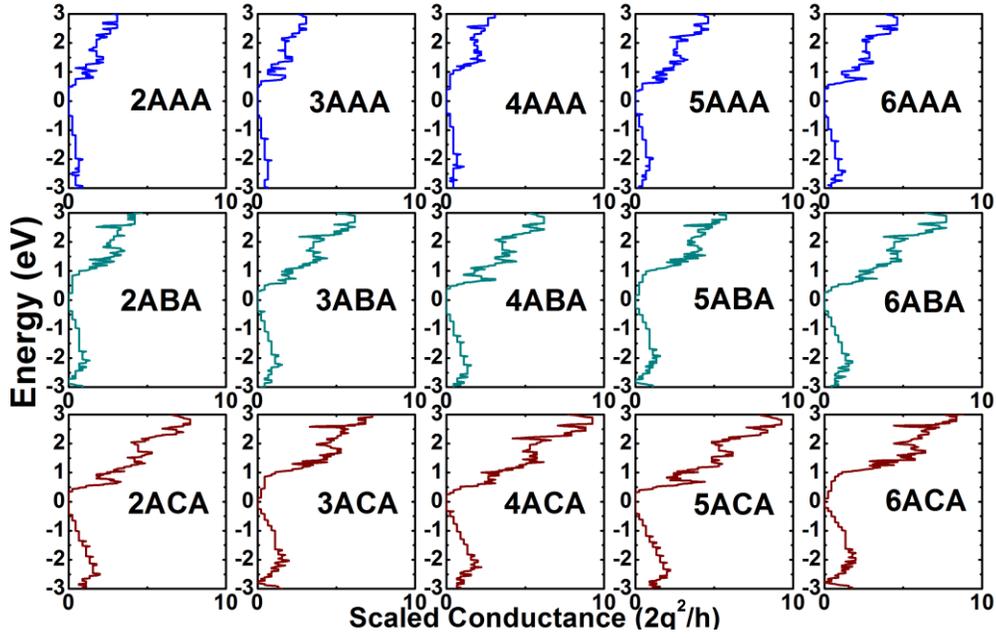

**Fig. 8. (b)** The scaled transmission spectra in zigzag direction for different stacking of layered black phosphorous.

For both electron and hole transport, in armchair and in zigzag directions, the ACA stacking shows the higher degree of channel transparency, followed by ABA and AAA stacking types.

It is worth mentioning here that as with the most state-of the art DFT-NEGF simulation frameworks [34-37] available, in the present work electron-phonon scattering has not been implemented. For applications such as next generation nanoscale FETs very short channel lengths (~10 nm and smaller) are being investigated. For such lengths it may be reasonable to assume fully ballistic transport in few layered 2D materials. The most prominent scattering mechanism in perfect sheets (defect free) of multilayer black phosphorous (and most of the layered semiconductors) is the electron-phonon scattering. [16] As a result there would be a likely suppression of transmission depending upon the temperature and the electron phonon coupling in the material. Still, the results presented in this work can give significant indications regarding fully ballistic transport in layered black phosphorous, which may be useful for analyzing the performance limits of black phosphorous based field effect transistors and other nanoelectronics devices.

Qualitatively it is possible to say the following, different stacking sequences, due to the difference in interlayer interaction strength would manifest themselves in variations in the elastic moduli and the deformation potentials. The 2D carrier mobility and therefore the quasi-ballistic transmission (inclusive of phonon scattering effects) would also depend on these two parameters.

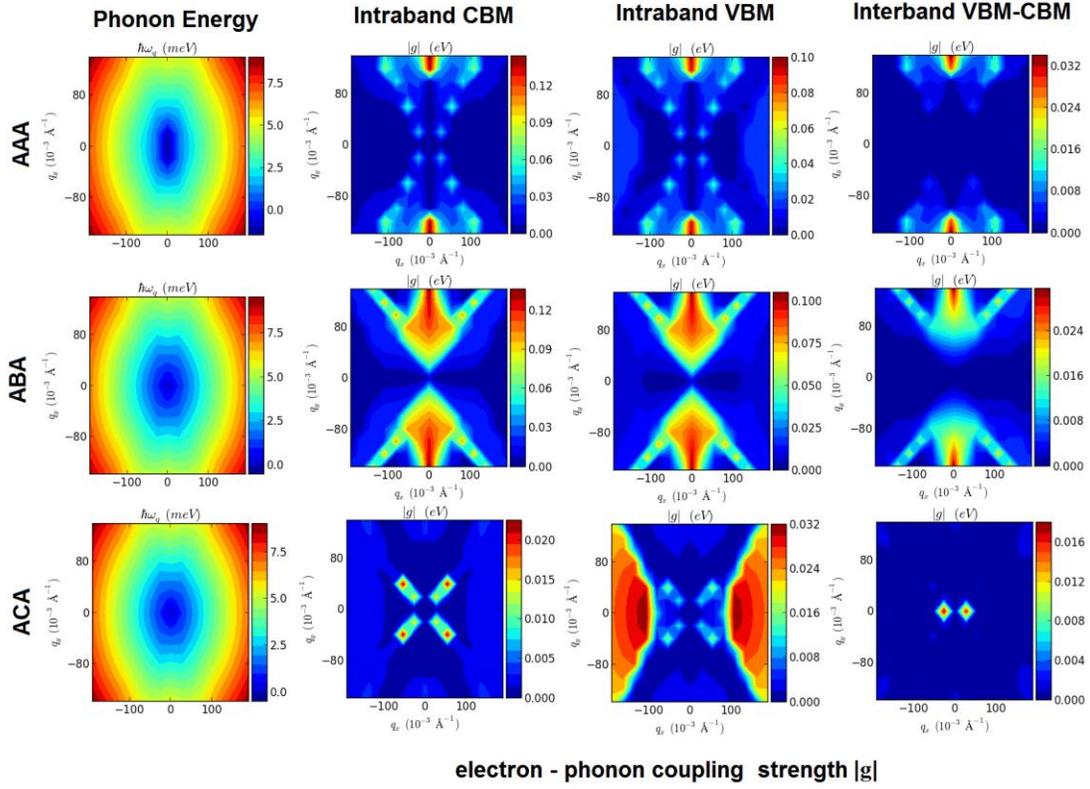

**Fig. 9 (a):** Contour plots of electron-phonon coupling strength for the transverse acoustic (TA) phonon modes for 3 layer black phosphorous stacks.

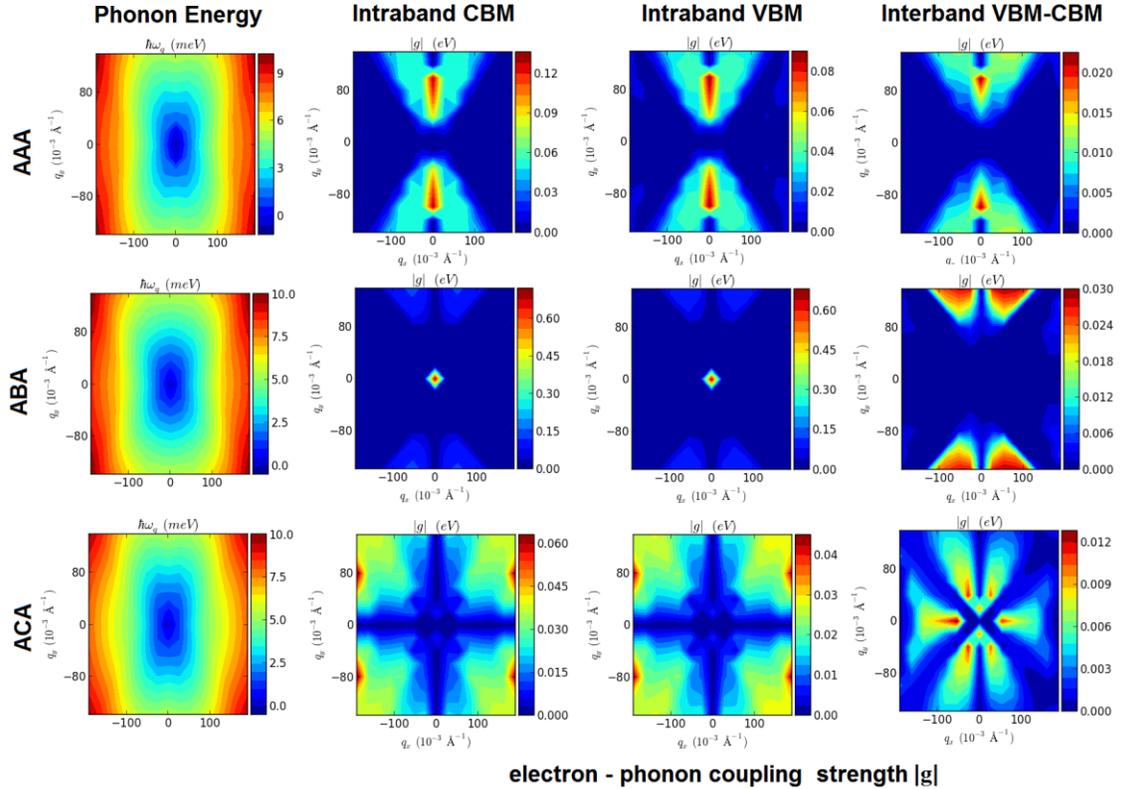

**Fig. 9 (b):** Contour plots of electron-phonon coupling strength for the longitudinal acoustic (LA) phonon modes for 3 layer black phosphorous stacks.

The electron-phonon coupling strengths $\left|g_{kq}^{\lambda}\right|$, in the three different types of stacks is shown in Fig. 9 (a) - (b). The intraband (for both VB maxima and CB minima) and the interband (VB to CB transition) coupling strengths are shown with a scaled coupling matrix with an energy broadening of 0.01 eV.

Here we have shown the zero order electron phonon coupling only for the transverse (TA) and longitudinal acoustic (LA) phonon modes, as they are the most important from the point of view of carrier transport. [16,38] Also in case of 2D channels in nanoscale devices the homopolar modes are likely to be quenched due to physical presence of the top gate and substrate (which constrain the channel in z direction). [38]

From the plots we can observe (note the difference in the color bar values) a significantly weaker intraband electron-phonon coupling (both VBM and CBM) in ACA stacks for both LA and TA phonon modes, as compared to those for the AAA and the ABA stacks. Among the other two types of stacks, ABA seems to have stronger coupling strengths than AAA for the TA phonon modes (for both VBM and CBM intraband transitions), while the case is just the reverse for the LA phonon modes. For interband (VBM to CBM transition) the coupling strengths however do not show that much amount of disparity in terms of magnitude (although spatially the contours change significantly).

The weaker electron-phonon coupling in ACA qualitatively implies that in such stacking, the carrier transport for both holes and electrons would experience lesser quenching compared to the AAA and the ABA stacked black phosphorous.

The optical absorption coefficient of the differently stacked layered black phosphorous is shown in Fig. 10. In general the absorption peaks are located in the energy range between 5 – 7 eV. The absorption plots (broadening of 0.1 eV applied) show a general trend of the increase in absorption and a slight sharpening of the peaks with the increase of the number of layers. The absorption peaks in near 5.2, 6.1 and 7 eV look to gain more strength as more layers of black phosphorous are added, irrespective of the nature of stacking. At the same time, there is a suppression of the smaller peaks in the range 5.5-5.7 eV, when the number of layers becomes more than four. This selective strengthening and weakening of absorption peaks, depending on addition of layers is more prominent in case of AAA and ACA stacks. For ACA, this preference for the peak near 5.2 eV for is most prominent.

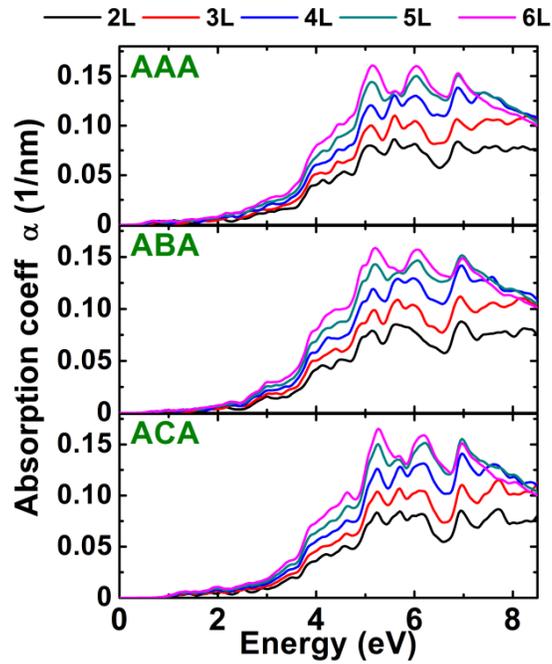

**Fig. 10.** The optical absorption spectra for different stacking sequences.

The absorption results although single particle, (without electron-hole interaction and excitonic corrections) can qualitatively predict the possibility of tailoring the optical absorption properties of black phosphorous with the number of layers and the stacking sequence.

## 4. Conclusion

In this paper we have studied the material and electrical properties of layered black phosphorous under different stacking schemes (namely AAA, ABA and ACA). For the said layered systems, our simulations showed a significant variation of bandstructure, carrier effective masses and carrier transmission for different stacking orders. The results indicate a high degree of tunability for these properties with the type of stacking chosen and the number of layers involved. This wide range of values for (1.1 eV-0.25 eV) bandgap variation, 2-3 folds reduction in hole effective mass in Zigzag direction, variations in optical absorption probabilities and significant change in carrier transmission spectra for combination of different stacking and number of layers can open up various possibilities to exploit layer engineering in black phosphorous for future nanoelectronics.

### Acknowledgement

The authors thank Dr. A. Blom and Dr. K. Stockbro of QuantumWise A/S for valuable discussions. This work was supported by DST, Govt. of India under DST INSPIRE Faculty Scheme, (IFA-13 ENG-62) and the DFG project HE 3543/26-1.

# Figure Captions:

**Fig. 1.** The optimized structures for layered black phosphorous showing the side and top views of the stacked unit-cell. Layers in site A , B and C are color coded for better understanding of the orientation from top.

**Fig. 2.** The bandstructure of monolayer black phosphorous calculated with different functionals.

**Fig. 3.** The formation energy per atom (obtained with GGA-PBE) of various stacked multilayer black phosphorous.

**Fig. 4.** The bandstructures of various stacked black phosphorous calculated with MGGA-TB09LDA.

**Fig. 5.** The variation of band-gap (calculated with MGGA-TB09LDA) of black phosphorus with the nature of stacking and the number of layers.

**Fig. 6.** The electron and hole effective masses at Γ point in the armchair and the zigzag directions for the differently stacked multilayer black phosphorous. (calculated with MGGA-TB09LDA)

**Fig. 7:** Spatial structure of wavefunctions of the VBM and the CBM states illustrated in the xy (top view) and yz (side view) planes using an isosurface of 0.06 e Å$^{-3}$.

**Fig. 8. (a)** The scaled transmission spectra in armchair direction for different stacking of layered black phosphorous.

**Fig. 8. (b)** The scaled transmission spectra in zigzag direction for different stacking of layered black phosphorous.

**Fig. 9 (a):** Contour plots of electron-phonon coupling strength for the transverse acoustic (TA) phonon modes for 3 layer black phosphorous stacks.

**Fig. 9 (b):** Contour plots of electron-phonon coupling strength for the longitudinal acoustic (LA) phonon modes for 3 layer black phosphorous stacks.

**Fig. 10.** The optical absorption spectra for different stacking sequences.